\documentclass[11pt]{article}
\usepackage{amssymb}
\usepackage{amsmath}
\usepackage{latexsym}
\catcode`@=11
\renewcommand{\section}{\@startsection{section}{1}{0pt}{\medskipamount}
{\medskipamount}{\large\bf}}
\numberwithin{equation}{section}
\catcode`@=12
\def\th{\theta}

\def\g{\gamma}
\def\de{\delta}

\def\o{\omega}

\def\vp{\varphi}
\def\vt{\vartheta}
\def\p{\phi}
\def\s{\sigma}

\def\sfrac#1#2{{\textstyle\frac{#1}{#2}}}
\def\m{\mu}
\def\n{\nu}
\def\pa{\partial}

\newcommand{\im}{\mathrm{i}}
\newcommand{\diff}{\mathrm{d}}
\newcommand{\rc}{{\mathbb{R}^{2n}}}

\newcommand{\rts}{{\mathbb{R}^{2n}_\theta{\times}S^2}}
\newcommand{\rt}{{\mathbb{R}^{2n}_\theta}}
\newcommand{\rs}{{\mathbb{R}^2{\times}S^2}}
\newcommand{\rns}{{\mathbb{R}^{2n}{\times}S^2}}
\newcommand{\R}{{\mathbb{R}}}

\newcommand{\Hcal}{{\cal H}}
\newcommand{\fh}{\hat{f}}

\newcommand{\xh}{\hat{x}}
\newcommand{\zh}{\hat{z}}
\newcommand{\zbh}{\hat{\bar{z}}}
\newcommand{\yb}{\bar{y}}
\newcommand{\zb}{\bar{z}}
\newcommand{\ab}{\bar{a}}
\newcommand{\bb}{\bar{b}}
\newcommand{\ca}{{\cal{A}}}
\newcommand{\cf}{{\cal{F}}}

\def\>{\rangle}
\def\<{\langle}
\def\+{\dagger}
\def\={\ =\ }

\begin{document}
\begin{titlepage}
\setcounter{page}{0}
\begin{flushright}
hep-th/0305195\\
ITP-UH-02/03
\end{flushright}

\vskip 2.5cm

\begin{center}

{\Large\bf Noncommutative Multi-Instantons on $\rns$}

\vspace{15mm}

{\Large Tatiana A. Ivanova${}^\+$} \ \ and \ \ {\Large Olaf Lechtenfeld${}^*$}
\\[10mm]
\noindent ${}^\+${\em Bogoliubov Laboratory of Theoretical Physics, JINR\\
141980 Dubna, Moscow Region, Russia}\\
{Email: ita@thsun1.jinr.ru}
\\[10mm]
\noindent ${}^*${\em Institut f\"ur Theoretische Physik,
Universit\"at Hannover \\
Appelstra\ss{}e 2, 30167 Hannover, Germany }\\
{Email: lechtenf@itp.uni-hannover.de}

\vspace{20mm}

\begin{abstract}
\noindent
Generalizing self-duality on $\R^2{\times}S^2$ to higher dimensions,
we consider the Donaldson-Uhlenbeck-Yau equations on $\R^{2n}{\times}S^2$
and their noncommutative deformation for the gauge group~U(2). 
Imposing SO(3) invariance (up to gauge transformations) reduces
these equations to vortex-type equations for an abelian gauge field and 
a complex scalar on~$\R^{2n}_\theta$. For a special $S^2$-radius~$R$ 
depending on the noncommutativity~$\theta$ we find explicit solutions in terms
of shift operators. These vortex-like configurations on~$\R^{2n}_\theta$ 
determine SO(3)-invariant multi-instantons on~$\R^{2n}_\theta{\times}S^2_R$
for $R=R(\theta)$. The latter may be interpreted as sub-branes 
of codimension~$2n$ inside a coincident pair of noncommutative D$p$-branes 
with an $S^2$ factor of suitable size.
\end{abstract}

\end{center}
\end{titlepage}

\section{Introduction}

Noncommutative deformation is a well established framework 
for stretching the limits of conventional (classical and quantum) 
field theories~\cite{Seiberg, Harvey}.
On the nonperturbative side, all celebrated classical field configurations
have been generalized to the noncommutative realm.
Of particular interest thereof are BPS~configurations, which are subject to 
first-order nonlinear equations. The latter descend from the $4d$ Yang-Mills 
(YM) self-duality equations and have given rise to 
instantons~\cite{Belavin}, monopoles~\cite{BPS} and vortices~\cite{Taubes},
among others. Their noncommutative counterparts were introduced in
\cite{NS}, \cite{GN} and \cite{vortex}, respectively, and have been studied
intensely for the past five years (see~\cite{Haman} for a recent review).

String/M theory embeds these efforts in a higher-dimensional context,
and so it is important to formulate BPS-type equations in more than 
four dimensions. In fact, noncommutative instantons in higher dimensions
and their brane interpretations have recently been considered 
in~\cite{Witten, Blum, Nekrasov}. 
Yet already 20 years ago, generalized self-duality equations for YM fields
in more than four dimensions were proposed~\cite{Corrigan, Ward} and their
solutions investigated e.g. in~\cite{Ward, Fairlie}.
For U$(k)$ gauge theory on a K\"ahler manifold these equations specialize
to the Donaldson-Uhlenbeck-Yau (DUY) equations~\cite{Donaldson, Uhlenbeck}.
They are the natural analogues of the $4d$ self-duality equations. 

In this letter we generalize the DUY equations to the noncommutative spaces
$\R^{2n}_{\th}{\times}S^2$ and construct explicit U$(2)$ multi-instanton 
solutions even though these equations are not integrable. The key lies in a 
clever ansatz for the gauge potential, due to Taubes~\cite{Taubes}, which 
we generalize to higher dimensions and to the noncommutative setting. 
This SO(3)-invariant ansatz reduces the U(2)~DUY~equations to vortex-type 
equations on $\R^{2n}_\th$. 
For $n{=}1$ the latter are the standard vortex equations on~$\R^2_\th$, 
while for $n{=}2$ they are intimately related to the Seiberg-Witten monopole 
equations on~$\R^4_\th$~\cite{PSW}.

\section{ Donaldson-Uhlenbeck-Yau equations on $\rts$}

\noindent
{\bf Manifold $\rts$.} 
We consider the manifold $\rns$ with the Riemannian metric
\begin{equation}\label{metric}
\diff s^2\= \sum_{\mu,\nu=1}^{2n}\de_{\mu\nu}\,\diff x^\mu \diff x^\nu
+ R^2 (\diff \vt^2 + \sin^2\vt\ \diff\vp^2)
\= \sum_{i,j=1}^{2n+2} g_{ij}\,\diff x^i \diff x^j\ ,
\end{equation}
where $x^1,\ldots, x^\mu, \ldots, x^{2n}$ are coordinates on $\rc$ while
$x^{2n+1}{=}\vt$ and $x^{2n+2}{=}\vp$ parametrize the standard two-sphere
$S^2$ with constant radius~$R$, i.e.~$0\le\vp\le 2\pi$ and $0\le\vt\le \pi$.
The volume two-form on $S^2$ reads
\begin{equation}\label{volume}
\sqrt{\det(g_{ij})}\ \diff\vt\wedge\diff\vp
\ =:\ \o_{\vt\vp}\ \diff\vt\wedge\diff\vp \= \o
\qquad\Longrightarrow\qquad
\o_{\vt\vp} = -\o_{\vp\vt}= R^2\sin\vt\ .
\end{equation}
The manifold $\rns$ is K\"ahler,
with local complex coordinates $z^1,\ldots,z^n,y$ where 
\begin{equation}\label{zz}
z^a\=x^{2a-1}-\im\,x^{2a} \qquad\textrm{and}\qquad
\zb^{\ab}\=x^{2a-1}+\im\,x^{2a} \qquad\textrm{with}\quad
a=1,\ldots,n
\end{equation}
and
\begin{equation}\label{zn1}
y\=\frac{R\,\sin\vt}{(1+\cos\vt )}\exp{(-\im\vp)}\quad ,\qquad
\yb\=\frac{R\,\sin\vt}{(1+\cos\vt) }\exp{(\im\vp)}\ ,
\end{equation}
so that $1{+}\cos\vt=\frac{2R^2}{R^2{+}y\yb}$.
In these coordinates, the metric takes the form\footnote{
{}From now on we use the Einstein summation convention for repeated indices.}
\begin{equation}
\diff s^2 \= \de_{a\bb}\ \diff z^a \diff\zb^{\bb}
+ \sfrac{4\,R^4}{(R^2+y\yb)^2}\,\diff y\,\diff\yb
\end{equation}
with $\de_{a\ab}{=}\de^{a\ab}{=}1$ (other entries vanish),
and the K\"ahler two-form reads
\begin{equation}\label{kahler}
\Omega\=-\sfrac{{\im}}{2}\bigl\{
\delta_{a\bb}\ \diff z^a\wedge\diff \zb^{\bb} +
\sfrac{4\,R^4}{(R^2+y\yb)^2}\ \diff y\wedge\diff \yb\bigr\}
\=-\sfrac{{\im}}{2}\de_{a\bb}\ \diff z^a\wedge\diff \zb^{\bb} +
\o_{\vt\vp}\ \diff \vt\wedge\diff \vp \ .
\end{equation}
For later use, we also note here the derivatives
\begin{equation}\label{pazz}
\pa_{z^a} \= \sfrac{1}{2}(\pa_{2a-1}+ \im \pa_{2a}) \qquad\textrm{and}\qquad
\pa_{\zb^{\ab}} \= \sfrac{1}{2}(\pa_ {2a-1} - \im \pa_{2a})\ ,
\end{equation}
where $\pa_{\m}\equiv {\pa}/\pa {x^\m}$ for $\mu{=}1,\ldots,2n$.

Classical field theory on the noncommutative deformation 
$\R^{2n}_\th$ of $\R^{2n}$ may be realized in a star-product formulation 
or in an operator formalism.
While the first approach alters the product of functions on~$\R^{2n}$
the second one turns these functions~$f$ into linear operators~$\fh$ 
acting on the $n$-harmonic-oscillator Fock space~$\Hcal$.
The noncommutative space~$\R^{2n}_\th$ may then be defined by declaring 
its coordinate functions $\xh^1,\ldots,\xh^{2n}$ to obey 
the Heisenberg algebra relations
\begin{equation}
[ \xh^\mu\,,\,\xh^\nu ] \= \im\,\th^{\mu\nu}
\end{equation}
with a constant antisymmetric tensor~$\th^{\mu\nu}$. 
The coordinates can be chosen in such a way that the matrix $(\th^{\m\n})$ 
will be block-diagonal with non-vanishing components
\begin{equation}\label{tha}
\th^{{2a-1}\ {2a}} \= -\th^{{2a}\ {2a-1}} \ =:\ \th^a \ .
\end{equation}
We assume that all $\th^a\ge0$; 
the general case does not hide additional complications.
For the noncommutative version of the complex coordinates (\ref{zz}) we have
\begin{equation}\label{zzb}
[\zh^a,\zbh^{\bb} ] \= -2\de^{a\bb}\,\th^a \ 
=:\ \th^{a\bb} \= -\th^{\bb a}\ \le 0
\quad,\qquad \mbox{and all other commutators vanish}\ .
\end{equation}
The Fock space~$\Hcal$ is spanned by the basis states
\begin{equation}
|k_1,k_2,\ldots,k_n\>\=\prod_{a=1}^{n}(2\th^a k_a!)^{-1/2}(\zh^{a})^{k_a} |0\>
\qquad \textrm{for} \quad k_a=0,1,2,\ldots \ ,
\end{equation}
which are connected by the action of creation and annihilation operators
subject to
\begin{equation}
\Bigl[\,\frac{\zbh^{\bb}}{\sqrt{2\th^b}}\ ,\ \frac{\zh^a}{\sqrt{2\th^a}}\,
\Bigr] \= \de^{a\bb} \ .
\end{equation}
We recall that, in the operator realization $f{\mapsto}\fh$,
derivatives of~$f$ get mapped according to
\begin{equation}\label{pazzf}
\pa_{z^a} f \ \mapsto\ \th_{a\bb}\,[\zbh^{\bb} , \fh] 
\ =:\ \pa_{\zh^a} \fh
\qquad\textrm{and}\qquad
\pa_{\zb^{\ab}} f \ \mapsto\ \th_{\ab b}\,[\zh^b , \fh]
\ =:\ \pa_{\zbh^{\ab}} \fh\ ,
\end{equation}
where $\th_{a\bb}$ is defined via $\th_{b\bar{c}}\th^{\bar{c}a}=\de^a_b$
so that $\th_{a\bb}=-\th_{\bb a}=\frac{\de_{a\bb}}{2\th^a}$.
Finally, we have to replace
\begin{equation}
\int_{\R^{2n}} \!\diff^n x\,f\ \ \mapsto\ \
\Bigl( \prod_{a=1}^n 2\pi\th^a \Bigr)\,\textrm{Tr}_\Hcal\,\fh\ .
\end{equation}

Tensoring $\R^{2n}_\th$ with a commutative $S^2$ means extending the
noncommutativity matrix $\th$ by vanishing entries in the two new directions.
A more detailed description of noncommutative field theories 
can be found in the review papers~\cite{Harvey}.

\bigskip

\noindent
{\bf Donaldson-Uhlenbeck-Yau equations.}
Let $M_{2q}$ be a complex $q{=}n{+}1$ dimensional K\"ahler manifold 
with some local real coordinates $x=(x^i)$ and a tangent space basis 
$\pa_{i} :=\pa /\pa x^i$ for $i,j =1,\ldots,2q$, so that a metric and
the K\"ahler two-form read $\diff s^2=g_{ij}\diff x^i\diff x^j$ and
$\Omega=\Omega_{ij}\,\diff x^i\wedge\diff x^j$, respectively.
Consider a rank~$k$ complex vector bundle over~$M_{2q}$ with 
a gauge potential ${\ca}={\ca}_{i}\diff x^i$ and 
the curvature two-form $\cf =\diff {\ca}+{\ca}\wedge {\ca}$ with components
${\cf}_{ij}=\pa_{i}{\ca}_{j}-\pa_{j}{\ca}_{i}+[{\ca}_{i},{\ca}_{j}]$.
Both ${\ca}_{i}$ and ${\cf}_{ij}$ take values in the Lie algebra $u(k)$.
The Donaldson-Uhlenbeck-Yau (DUY) equations~\cite{Donaldson, Uhlenbeck} 
on $M_{2q}$ are
\begin{equation}\label{DUY}
*\Omega\wedge {\cf}\ =\ 0 \qquad\textrm{and}\qquad {\cf}^{0,2}\=0\ ,
\end{equation}
where $\Omega$ is the K\"ahler two-form, ${\cf}^{0,2}$ is the $(0,2)$ part of 
${\cf}$, and $*$ is the Hodge operator. 
In our local coordinates $(x^i)$ we have
$q!(*\Omega\wedge{\cf})=(\Omega,{\cf})\Omega^q=\Omega^{ij}{\cf}_{ij}\Omega^q$
where $\Omega^{ij}$ are defined via $\Omega^{ij}\Omega_{jk}=\de^i_k$.
Due to the antihermiticity of~$\cf$, it follows that also ${\cf}^{2,0}=0$.
For $q{=}2$ the DUY equations (\ref{DUY}) coincide with the anti-self-dual 
Yang-Mills (ASDYM) equations 
\begin{equation}\label{sdym1}
* {\cf} \ =\ - {\cf} 
\end{equation}
introduced in~\cite{Belavin}.

Specializing now $M_{2q}$ to be $\rns$, the DUY equations (\ref{DUY})
in the local complex coordinates $(z^a,y)$ take the form
\begin{equation}\label{DU}
\de^{a\bb}{\cf}_{z^a\zb^{\bb}}+\sfrac{(R^2+y\yb)^2}{4\,R^4} {\cf}_{y\yb}\=0
\quad,\qquad 
{\cf}_{\zb^{\ab}\zb^{\bb}}\=0 \qquad\mbox{and}\qquad {\cf}_{\zb^{\ab}\yb}\=0\ ,
\end{equation}
where $a,b=1,\ldots,n$.
Using formulae (\ref{zn1}), we obtain
\begin{align}\label{complfun}
{\cf}_{\zb^{\ab}\yb}&\={\cf}_{\zb^{\ab}\vt}\frac{\pa\vt}{\pa\yb}
                       +{\cf}_{\zb^{\ab}\vp}\frac{\pa\vp}{\pa\yb}
\=\frac{1}{\yb}(\sin\vt\, {\cf}_{\zb^{\ab}\vt}-\im\,{\cf}_{\zb^{\ab}\vp}) \ ,\\
{\cf}_{y\yb}&\={\cf}_{\vt\vp}\,\Bigl|\frac{\pa(\vt,\vp)}{\pa(y,\yb)}\Bigr|
\=\frac{1}{2\im}\,\frac{\sin\vt}{y\yb}\,{\cf}_{\vt\vp}
\=\frac{1}{2\im}\,\frac{(1{+}\cos\vt)^2}{R^2\sin\vt}\,{\cf}_{\vt\vp}
\end{align} 
and finally write the Donaldson-Uhlenbeck-Yau equations on $\rns$ 
in the alternative form
\begin{equation}\label{D}
2\im\ \delta^{a\bb}{\cf}_{z^a\zb^{\bb}}+\sfrac{1}{R^2\sin\vt}{\cf}_{\vt\vp}\=0
\quad,\qquad {\cf}_{\zb^{\ab}\zb^{\bb}}\=0 \quad,\qquad 
\sin\vt\, {\cf}_{\zb^{\ab}\vt}-\im\,{\cf}_{\zb^{\ab}\vp}\=0 \ .
\end{equation}

The transition to the noncommutative DUY equations is trivially achieved 
by going over to operator-valued objects everywhere. In particular,
the field strength components in~(\ref{D}) then read 
$\hat{\cf}_{ij}=\pa_{\xh^i}\hat{\ca}_{j}-\pa_{\xh^j}\hat{\ca}_{i}
+[\hat{\ca}_{i},\hat{\ca}_{j}]$, 
where e.g. $\hat{\ca}_{i}$ are simultaneously $u(k)$ and operator valued.
To avoid a cluttered notation, we drop the hats from now on.

\section{Generalized vortex equations on $\rt$}

\noindent
{\bf Noncommutative generalization of Taubes' ansatz.}
Considering the particular case (\ref{sdym1}) of the SU(2) DUY equations 
on $\rs$, Taubes introduced an $SO(3)$-invariant ansatz\footnote{
Similarly, Witten's ansatz~\cite{Witt} for gauge fields on $\R^4$ 
reduces (\ref{sdym1}) to the vortex equations on the hyperbolic space $H^2$ 
(cf.~\cite{Correa} for the noncommutative $\R^4$).}
for the gauge potential $\ca$ which reduces the ASDYM equations~(\ref{sdym1}) 
to the vortex equations on $\R^2$~\cite{Taubes} (see also~\cite{Forgacs}). 
Here we extend Taubes' ansatz to the higher-dimensional manifold $\rns$ 
and reduce the noncommutative\footnote{
As it is well known~\cite{Harvey}, in the noncommutative 
case one should use U(2) instead of SU(2).}
U(2) Donaldson-Uhlenbeck-Yau equations (\ref{D}) to generalized vortex 
equations on $\rt$, including their commutative ($\th{=}0$) limit.
In section 4, we will write down explicit solutions of the generalized 
noncommutative vortex equations on $\rc$ which determine multi-instanton 
solutions of the noncommutative YM equations on $\rns$. 

We begin with the $u(2)$-valued operator one-form $\ca$ on $\rts$.
Imposing SO(3)~invariance up to a gauge transformation, Taubes~\cite{Taubes} 
found for $n{=}1$ and $\th{=}0$ that the $S^2$ dependence of~$\ca$ 
must be collected in the $su(2)$~matrix 
\begin{equation}\label{Q}
Q\ =\ \im \begin{pmatrix}
\cos\vt & \textrm{e}^{-\im\vp}\sin\vt \\ \textrm{e}^{\im\vp}\sin\vt & -\cos\vt
\end{pmatrix} \=
\im\ (\sin\vt\cos\vp\ \s_1 + \sin\vt\sin\vp\ \s_2 + \cos\vt\ \s_3 )
\end{equation}
and its differential $\diff Q$.
Note that $Q^2=-1$ and $\frac{\pa Q}{\pa\vp}=-\sin\vt\,Q\,\frac{\pa Q}{\pa\vt}$. 
Our slight generalization of his ansatz to $\rts$ reads
(${\bf1}=(\begin{smallmatrix}1&0\\0&1\end{smallmatrix})$)
\begin{equation}\label{A}
{\ca} \= \sfrac{1}{2}\bigl\{ (\im\,Q - \g\,{\bf1}) A \,+\, 
(\p_1{-}1)Q\diff Q \,+\, \p_2\,\diff Q\bigr\}\ ,
\end{equation}
where the constant $\g$ parametrizes the additional $u(1)$ piece.
The one-form $A=A_\m(x)\diff x^\m$ with $A_\m\in u(1)\cong\im\R$ 
and $\mu{=}1,\ldots,2n$ is antihermitian 
while $\p_{1,2}=\p_{1,2}(x)\in\R$ are hermitian, 
all being operators in~$\Hcal$ only. 
Note that this form reduces the nonabelian connection~$\ca$ to the
abelian objects $(A,\p_1,\p_2)$ whose noncommutative character thus
does not interfere with the $u(2)$ structure.
Calculation of the curvature 
\begin{equation}\label{F}
{\cf} =\diff {\ca} + {\ca} \wedge {\ca} 
={\sfrac{1}{2}}{{\cf}_{ij}} \diff x^i\wedge\diff x^j = 
\sfrac{1}{2}{\cf}_{\m\n}\diff x^\m\wedge\diff x^\n +
{\cf}_{\m\vt}\diff x^\m\wedge\diff\vt + {\cf}_{\m\vp}\diff x^\m\wedge\diff\vp
+ {\cf}_{\vt\vp}\diff\vt\wedge\diff\vp
\end{equation}
for $\ca$ of the form (\ref{A}) yields
\begin{align}\label{Fmn}
2{\cf}_{\m\n} &= \im\,Q\bigl(\pa_{\m}A_{\n}{-}\pa_{\n}A_{\m}{-}
\g [A_{\m},A_{\n}]\bigr) -  \g{\bf1}\bigl(\pa_{\m}A_{\n}{-}\pa_{\n}A_{\m}{-}
\sfrac{1{+}\g^2}{2\g} [ A_{\m}, A_{\n}]\bigr)\ ,\\
\label{Fmvt}
4{\cf}_{\m\vt} &= \Bigl\{\! Q\bigl(2\pa_{\m}\p_{1}{+} \im A_{\m}\p_{2}{+}
\im\p_{2} A_{\m} {-} \g [A_{\m},\p_{1}]\bigr)+ {\bf1}\bigl( 2\pa_{\m}\p_{2}{-} 
\im A_{\m}\p_{1}{-}\im\p_{1}A_{\m} {-} \g [A_{\m},\p_{2}] \bigr) \!\Bigr\}
\frac{\pa Q}{\pa\vt}\ ,\\
\label{Fmvp}
4{\cf}_{\m\vp} &= \Bigl\{\! Q\bigl(2\pa_{\m}\p_{1}{+} \im A_{\m}\p_{2}{+}
\im\p_{2}A_{\m} {-} \g [A_{\m},\p_{1}]\bigr)+ {\bf1}\bigl( 2\pa_{\m}\p_{2}{-} 
\im A_{\m}\p_{1}{-} \im\p_{1}A_{\m} {-} \g [A_{\m},\p_{2}] \bigr) \!\Bigr\}
\frac{\pa Q}{\pa\vp}\ ,\\
\label{Fvtvp}
2{\cf}_{\vt\vp} &= \Bigl\{ 
Q\,\bigl(1{-}\p_1^2{-}\p_2^2\bigr)+{\bf1}\,[\p_1,\p_2] \Bigr\} \sin\vt \ .
\end{align}

In the complex coordinates (\ref{zz}) with 
$A_{z^a}=\sfrac{1}{2}(A_{2a-1}+\im\, A_{2a})$ and
$A^\+_{\zb^{\ab}}=-A_{z^a}$ we have
\begin{equation}\label{Faa}
{\cf}_{2a-1\ 2a}= - Q\bigl( \pa_{z^a}A_{\zb^{\ab}}-\pa_{\zb^{\ab}}A_{z^a}-
\g [A_{z^a},A_{\zb^{\ab}}]\bigr) - \im \g{\bf1}\bigl( \pa_{z^a}A_{\zb^{\ab}}-
\pa_{\zb^{\ab}}A_{z^a}-\sfrac{1{+}\g^2}{2\g} [A_{z^a},A_{\zb^{\ab}}]\bigr)
\end{equation}
which agrees with $2\im\,{\cf}_{z^a\zb^{\ab} }$.

\goodbreak
\bigskip

\noindent
{\bf Vortex-type equations in $\rt$.}\   
Introducing $\p : = \p_1 + \im \p_2$ and substituting (\ref{Fvtvp}) 
and (\ref{Faa}) into the first equation from (\ref{D}), we obtain 
$$
- \de^{a\bb}\Bigl\{Q\bigl( \pa_{z^a}A_{\zb^{\bb}}-\pa_{\zb^{\bb}}A_{z^a}- 
\g [A_{z^a},A_{\zb^{\bb}}]\bigr)+ \im \g{\bf1}\bigl( \pa_{z^a}A_{\zb^{\bb}}-
\pa_{\zb^{\bb}}A_{z^a}- \sfrac{1{+}\g^2}{2\g} [A_{z^a},A_{\zb^{\bb}}]\bigr)
\Bigl\}\ + \phantom{XXXX}
$$
\begin{equation}\label{D1}
\phantom{XXXXXXXXXXXXXX}
+\frac{1}{4R^2}\bigl(Q\,(2{-}\p\p^\+{-}\p^\+\p )+\im{\bf1}\,[\p ,\p^\+]\bigr)\=0
\end{equation}
which splits into the two equations
\begin{align}\label{D2}
\de^{a\bb} &\Bigl\{ \pa_{z^a}A_{\zb^{\bb}}-\pa_{\zb^{\bb}}A_{z^a} - 
\g [A_{z^a},A_{\zb^{\bb}}]\Bigr\}\= 
\frac{1}{4R^2}\bigl(2-\p\p^\+-\p^\+\p \bigr)\ , \\[6pt]
\label{D3}
\g\,\de^{a\bb} &\Bigl\{ \pa_{z^a}A_{\zb^{\bb}}-\pa_{\zb^{\bb}}A_{z^a} -
\sfrac{1{+}\g^2}{2\g} [A_{z^a},A_{\zb^{\bb}}]\Bigr\} \=
\frac{1}{4R^2}\,[\p,\p^\+] 
\end{align}
after separating into the $su(2)$ (proportional to~$Q$) and $u(1)$
(proportional to $\im{\bf1}$) components.

The second equation from (\ref{D}) can be written as
\begin{equation}\label{D4}
Q\bigl( \pa_{\zb^{\ab}}A_{\zb^{\bb}}-\pa_{\zb^{\bb}}A_{\zb^{\ab}} - 
\g [A_{\zb^{\ab}},A_{\zb^{\bb}}]\bigr)
+\im \g{\bf1}\bigl( \pa_{\zb^{\ab}}A_{\zb^{\bb}}-\pa_{\zb^{\bb}}A_{\zb^{\ab}}- 
\sfrac{1{+}\g^2}{2\g} [A_{\zb^{\ab}},A_{\zb^{\bb}}]\bigr)\=0\ .
\end{equation}
After some algebra, using (\ref{Fmvt}) and (\ref{Fmvp}), 
we find that the third equation from (\ref{D}) is equivalent to
\begin{equation}\label{D5}
2\pa_{\zb^{\ab}}\p + (1{-}\g) A_{\zb^{\ab}}\p + (1{+}\g)\p A_{\zb^{\ab}}\=0\ .
\end{equation}

Let us consider the commutative case $\th^{\m\n}=0$ and put $\g =0$. Then
the Donaldson-Uhlenbeck-Yau equations on $\rns$ for $\ca$ defined in (\ref{A}) 
reduce to
\begin{align}\label{vort1}
\de^{a\bb}\Bigl\{ \pa_{z^a}A_{\zb^{\bb}}-\pa_{\zb^{\bb}}A_{z^a}\Bigr\} &\=  
\frac{1}{2R^2}\bigl(1-\p\bar{\p} \bigr)\ , \\
\label{vort2}
\pa_{\zb^{\ab}}A_{\zb^{\bb}}-\pa_{\zb^{\bb}}A_{\zb^{\ab}} &\=0\ ,\\[4pt]
\label{vort3}
\pa_{\zb^{\ab}}\p + A_{\zb^{\ab}}\p &\= 0\ ,
\end{align}
where $\bar{\p}$ is the complex conjugate of the scalar field $\p$.
Equations (\ref{vort1})--(\ref{vort3}) generalize the vortex 
equations~\cite{Taubes} on $\R^2$ to the higher-dimensional space $\rc$.

For the noncommutative case $\th^{\m\n}\ne 0$ we choose $\g =-1$.
Comparing (\ref{D2}) and (\ref{D3}), 
we obtain a constraint equation on the field $\p$,
\begin{equation}\label{cons}
2-\p\p^\+ -\p^\+\p\=-\,[\p,\p^\+]
\qquad\Longrightarrow\qquad \p^\+\p\=1\ ,
\end{equation}
and the following noncommutative generalization of the vortex equations 
in $2n$ dimensions:
\begin{align}\label{nvor1}
\de^{a\bb}\,F_{z^a\zb^{\bb}}\ :=\
\de^{a\bb}\Bigl\{ \pa_{z^a}A_{\zb^{\bb}}-\pa_{\zb^{\bb}}A_{z^a} + 
[A_{z^a},A_{\zb^{\bb}}]\Bigr\} &\= \frac{1}{4R^2}\Bigl(1-\p\p^\+ \Bigr)\ , \\
\label{nvor2}
F_{\zb^{\ab}\zb^{\bb}}\ :=\ 
\pa_{\zb^{\ab}}A_{\zb^{\bb}}-\pa_{\zb^{\bb}}A_{\zb^{\ab}} + 
[A_{\zb^{\ab}},A_{\zb^{\bb}}] &\=0\ ,\\[4pt]
\label{nvor3}
\pa_{\zb^{\ab}}\p + A_{\zb^{\ab}}\p &\= 0\ .
\end{align}
These equations and their antecedent DUY equations on $\rts$ are not integrable
even for $n{=}1$.
Therefore, neither dressing nor splitting approaches, developed in~\cite{LP1} 
for integrable equations on noncommutative spaces, can be applied. 
The modified ADHM construction~\cite{NS} also does not work in this case. 
However, some special solutions can be obtained by choosing a proper ansatz
as we shall see next.

\section{Multi-instanton solutions on $\rts$}

\noindent
{\bf Solutions of the constrained vortex-type equations.}
We are going to present explicit solutions to the noncommutative generalized
vortex equations (\ref{nvor1}) -- (\ref{nvor3}) 
subject to the constraint~(\ref{cons}). The latter can be solved by putting
\begin{equation}\label{phi}
\p \= S_N
\qquad\mbox{and}\qquad
\p^\+ \= S_N^\+ \ ,
\end{equation}
where $S_N$ is an order-$N$ shift operator acting on the Fock space~$\cal H$,
i.e.
\begin{equation}\label{phiprop}
S_N^\+S_N \=1 \qquad\mbox{while}\qquad
S_N S_N^\+ \=1 - P_N\ ,
\end{equation}
with $P_N$ being a hermitean rank-$N$ projector: $P_N^2= P_N= P^{\+}_N$. 

It is convenient to introduce the operators 
\begin{equation}\label{X}
X_{z^a} \= A_{z^a} + \th_{a\bb}\,\zb^{\bb} 
\qquad\mbox{and}\qquad 
X_{\zb^{\ab}} \= A_{\zb^{\ab}} + \th_{\ab b}\,z^b
\end{equation}
in terms of which
\begin{equation}
F_{z^{a}\zb^{\bb}}\= [X_{z^a},X_{\zb^{\bb}}] + \th_{a\bb}
\qquad\textrm{and}\qquad
F_{\zb^{\ab}\zb^{\bb}}\= [X_{\zb^{\ab}},X_{\zb^{\bb}}] \ .
\end{equation}
We now employ the shift-operator ansatz (see e.g.~\cite{GN,Strom})
\begin{equation}\label{ansatz}
X_{z^a} \=\th_{a\bb}\,S_N\,\zb^{\bb}\,S_N^\+ 
\qquad\mbox{and}\qquad 
X_{\zb^{\ab}} \=\th_{\ab b}\,S_N\,z^{b}\,S_N^\+
\end{equation}
for which
\begin{equation}\label{ansatzF}
F_{z^{a}\zb^{\bb}} \= \th_{a\bb}\,P_N \= \de_{a\bb}\,\frac{P_N}{2\th^a} 
\qquad\textrm{and}\qquad F_{\zb^{\ab}\zb^{\bb}} \= 0
\end{equation}
since $\th_{a\bb}=\frac{\de_{a\bb}}{2\th^a}$.
After substituting (\ref{phi}) and (\ref{ansatzF}) 
into the first vortex equation~(\ref{nvor1}), 
we obtain the condition
\begin{equation}\label{res}
\de^{a\bb}\th_{a\bb}\,P_N \= \frac{1}{4R^2} P_N 
\qquad\Longleftrightarrow\qquad   
\frac{1}{\th^1}+\ldots+ \frac{1}{\th^n} \= \frac{1}{2R^2}\ .    
\end{equation}
The remaining vortex equations (\ref{nvor2}) and (\ref{nvor3}) 
are identically satisfied by (\ref{phi}) and~(\ref{ansatzF}).

Hence, for $\gamma=-1$ we have established on $\R^{2n}$ a whole class
of noncommutative constrained vortex-type configurations
\begin{equation}
A_{z^a} \= \th_{a\bb}\,\bigl( S_N\,\zb^{\bb}\,S_N^\+ - \zb^{\bb} \bigr)
\qquad\mbox{and}\qquad \p \= S_N \ ,
\end{equation}
parametrized by shift operators~$S_N$.
Our particular form~(\ref{A}) for~$\ca$ then yields a plethora of solutions 
to the noncommutative DUY equations on~$\rns$.
These configurations generalize U(2) multi-instantons from $\rs$ to $\rts$.
To substantiate this interpretation we finally calculate 
their topological charge.

\goodbreak
\bigskip

\noindent
{\bf Topological charge.}
For $\g=-1$, from (\ref{Fvtvp}) and (\ref{Faa}) we get
\begin{equation}
{\cf}_{\vt\vp} \= \sfrac{1}{4}(Q-\im{\bf1})\,\sin\vt\ P_N 
\qquad\mbox{and}\qquad
{\cf}_{2a-1\ 2a} \= (\im{\bf1}-Q)F_{z^{a}\zb^{\ab}} \= 
(Q-\im{\bf1})\,\frac{P_N}{2\th^a} \ .
\end{equation}
Employing
\begin{equation}
(Q-\im{\bf1})^{n+1}\=(-2\im)^n(Q-\im{\bf1})
\qquad\mbox{and}\qquad 
\mbox{tr}_{2\times 2} (Q-\im{\bf1}) \= -2\im
\end{equation}
we have
\begin{align}
\mbox{tr}_{2\times 2}\underbrace{{\cf}\wedge\ldots\wedge {\cf}}_{n+1} &\= 
(n{+}1)!\ \mbox{tr}_{2\times 2}
{\cf}_{12}{\cf}_{34}\ldots{\cf}_{2n-1\,2n} {\cf}_{\vt\vp}\
\diff x^1\wedge\diff{x^2}\wedge\ldots\wedge\diff x^{2n}
\wedge\diff\vt\wedge\diff\vp \nonumber\\
&\= (n{+}1)!\ \frac{(-2\im)^{n+1}}{2^{n+2}}\ \frac{P_N}{\prod_{a=1}^n{\th^a}}\
\diff x^1\wedge\diff{x^2}\wedge\ldots\wedge\diff x^{2n}
\wedge\sin\vt\,\diff\vt\wedge\diff\vp\ .
\end{align}
With this, the topological charge indeed becomes
\begin{align}
{\cal Q}\ :=\ &\frac{1}{(n{+}1)!}\ \Bigl(\frac{\im}{2\pi}\Bigr)^{n+1}
\Bigl(\prod_{a=1}^n{2\pi\th^a}\Bigr)\,\mbox{Tr}_{\cal H}
\int_{S^2} \mbox{tr}_{2\times 2}
\underbrace{{\cf}\wedge\ldots\wedge {\cf}}_{n+1} \nonumber\\
\= &\Bigl(\frac{\im}{2\pi}\Bigr)^{n+1}\,\frac{(-2\im)^{n+1}}{2^{n+2}}\, 
\Bigl(\prod_{a=1}^n{2\pi\th^a}\Bigr)
\Bigl(\mbox{Tr}_{\cal H}\frac{P_N}{\prod_{a=1}^n{\th^a}}\Bigr)\
\int\limits_{S^2} \sin\vt\,\diff\vt\wedge\diff\vp \nonumber\\
\= &\frac{1}{4\pi}\,\bigl(\mbox{Tr}_{\cal H} P_N\bigr)\
\int\limits_{S^2} \sin\vt\,\diff\vt\wedge\diff\vp \= N\ .
\end{align}

\section{Concluding remarks}

\noindent
By solving the noncommutative Donaldson-Uhlenbeck-Yau equations 
we have presented explicit U(2) multi-instantons on $\rts$ which are 
uniquely determined by abelian vortex-type configurations on~$\rt$. 
The existence of these solutions required the condition~(\ref{res}) 
relating the $S^2$-radius~$R$ to~$\th$ via
$\ R=(2\sum_{a=1}^n \frac{1}{\th^a})^{-1/2}$.
We see that any commutative limit ($\th^a{\to}0$) forces $R\to0$ as well,
and the configuration becomes localized in~$\rc$ (for $n{=}1$) or
disappears (for $n{>}1$).
The moduli space of our $N$-instanton solutions is that of rank-$N$ projectors
in the $n$-oscillator Fock space.

Since standard instantons localize all compact coordinates in the ambient space
they have been interpreted as sub-branes inside D$p$-branes
\cite{Seiberg,Harvey,Haman,Witten,Blum,Nekrasov}. The presence 
of an NS background $B$-field deforms such configurations noncommutatively.
In the same vein, the solutions presented in this letter may be viewed as
a collection of $N$ sub-branes of codimension~$2n$, i.e.~as D($p{-}2n$)-branes
located inside two coincident D$p$-branes, with all branes sharing a common 
two-sphere $S^2_R$.

\bigskip

\noindent
{\bf Acknowledgements.}
The authors are grateful to A.D.~Popov for fruitful discussions and
for reading the manuscript. O.L. wishes to thank B.-H.~Lee for discussions.
T.A.I. acknowledges the Heisenberg-Landau Program for partial support and
the Institut f\"ur Theoretische Physik der Universit\"at Hannover for 
its hospitality. This work was partially supported by grant LE-838/7 within 
the framework of the DFG priority program (SPP 1096) in string theory.


\end{document}